\documentclass[journal=jpcbfk,manuscript=article]{achemso}
\usepackage[version=3]{mhchem} % Formula subscripts using \ce{}
\usepackage[T1]{fontenc}       % Use modern font encodings
\usepackage[utf8]{inputenc}

\newcommand{\angstrom}{\mbox{\normalfont\AA}}

\author{Frédéric Poitevin}
\affiliation[IPhT, CEA]{Institut de Physique Théorique, Université Paris Saclay, CEA, UMR3681 du CNRS, Gif-sur-Yvette, France}
\altaffiliation{Present address: Department of Structural Biology, Stanford, CA 94305, USA}
\email{fpoitevi@stanford.edu}
\author{Marc Delarue}
\affiliation[DSM, IP]{Unit of Structural Dynamics of Macromolecules, UMR 3528 du CNRS, Institut Pasteur, 75015 Paris, France}
\author{Henri Orland}
\affiliation[IPhT, CEA]
{Institut de Physique Théorique, Université Paris Saclay, CEA, UMR3681 du CNRS, Gif-sur-Yvette, France}
\altaffiliation{Beijing Computational Science Research Center, Haidian District, Beijing 100094, China
}
\title[Sampling of electrolyte density fluctuations]
  {Beyond Poisson-Boltzmann: Numerical sampling of charge density fluctuations}

\abbreviations{SPDE}

\begin{document}

%%%%%%%%%%%%%%%%%%%%%%%%%%%%%%%%%%%%%%%%%%%%%%%%%%
\begin{abstract}
We present a method aimed at sampling charge density fluctuations in Coulomb systems. The derivation follows from a functional integral representation of the partition function in terms of charge density fluctuations. Starting from the mean-field solution given by the Poisson-Boltzmann equation, an original approach is proposed to numerically sample fluctuations around it, through the propagation of a Langevin-like stochastic partial differential equation (SPDE). The diffusion tensor of the SPDE can be chosen so as to avoid the numerical complexity linked to long-range Coulomb interactions, effectively rendering the theory completely local. A finite-volume implementation of the SPDE is described, and the approach is illustrated with preliminary results on the study of a system made of two like-charge ions immersed in a bath of counter-ions.
\end{abstract}

%%%%%%%%%%%%%%%%%%%%%%%%%%%%%%%%%%%%%%%%%%%%%%%%%%
\section{Introduction}
The description of charged, fluctuating systems is an old problem in chemistry and physics\cite{Onsager:1934ep,Kirkwood:1954cw}, with many applications across a variety of fields such as material science or structural molecular biology. In the late 1990s, one of us reintroduced a field-theoretic  formulation to tackle such systems\cite{Attard:1988fh,Podgornik:1988ij,BenTal:1994hk,Netz:1999vd,Borukhov:2000gk,Netz:2000kh,Netz:2003fw} that has since been widely used. This formulation has been either applied to the most accurate description of systems with simple geometries\cite{Moreira:1999tp,Netz:2000kh,Netz:2003fw,Naji:2005ul,Buyukdagli:2009dt,Levy:2012gt,Naji:2013gr,Wang:2013cla,Buyukdagli:2013uj,Buyukdagli:2014vr,Xu:2013tx,Ma:2014ij}, or to the approximated description of complex systems of arbitrary shape\cite{Azuara:2008hb,Koehl:2010fi,Buyukdagli:2016uk}. In recent years, much effort has been spent in the development of methods for dealing with such complex systems, namely "real" molecular objects, in a less approximate way. This article belongs to that effort.\\
We will be concerned with the distribution of freely moving ions around charged objects in some high-dielectric solvent. Due to the long-range nature of the Coulomb interaction, one ion is likely to interact with many different counter ions for a given configuration of the system. This particular feature makes the mean-field approach a very successful one in many cases. A mean-field description of the system can be carried out by solving the so-called Poisson-Boltzmann (PB) equation and can be used to describe experiments or simulations quantitatively. There are however several factors which contribute to deviations from the PB equation, including additional, short-range potentials (such as hardcore interactions) and solvent effects. In this article, as in Ref.\citenum{Netz:2000kh}, we consider deviations due to counter-ion fluctuations and use the field-theoretic framework to include such effects systematically. However, instead of including them in a loop-wise expansion, we include them through a numerical sampling approach\cite{Negele:1988uh}. This paper is organized as follows.\\
We first introduce the theory, by constructing the partition function for a simple system made of ions moving in a uniform dielectric medium surrounding a solvent-excluded volume containing fixed arbitrary charges. After briefly relating two possible approaches to deal with such systems, namely a density-functional description of the system's Hamiltonian, or a description in terms of conjugated quantities, namely potentials, we present a description of the system in terms of charge densities where the fluctuations around the PB solution are sampled by using a new type of Langevin stochastic partial differential equation (SPDE). The  novelty of this Langevin approach is that despite the long-range Coulomb interactions, it is completely local and involves at most second derivatives.\\
The second part of this paper deals with the numerical implementation of the method. Specifically, we describe how the SPDE is discretized, both in time and space. We are interested here in solving the electrostatics problem for systems of arbitrary 3D geometries. We aim at distributing this method as part of our Generalized Poisson-Boltzmann Equation solver, AquaSol\cite{Koehl:2010fi}, in the near future.\\
The final part of this paper presents preliminary results on the study of a simple system made of two like-charged ions immersed in a box containing counter-ions. The relative contribution of the mean-field Coulomb repulsion force and that of the fluctuation component are compared, suggesting like-charge attraction at intermediate distances as the valence of the ions increases. We finally discuss some possible difficulties of the current implementation, and highlight advantages the proposed method bears for the study of arbitrary Coulomb systems.

\section{Theory}
Consider a system made of $J$ types of mobile ions of type $\{\alpha=1,\ldots,J\}$, with valence $q_{\alpha}$, at positions $\mathbf{R}_{k}^{(\alpha)}$, in number $k=\{1,\ldots,N_{\alpha}\}$, and a solute made of impenetrable fixed charges, in a medium of uniform permittivity $\epsilon$, all interacting through the Coulomb potential. We define the number density operator, charge density operator for each species and the total charge density operator by
\begin{eqnarray}\label{defoperators}
\hat{n}_{\alpha}(\mathbf{r})&=&\sum_{k=1}^{N_{\alpha}}\delta(\mathbf{r}-\mathbf{R}_{k}^{(\alpha)})\\\nonumber
\hat{\rho}_{\alpha}(\mathbf{r})&=&q_{\alpha}\hat{n}_{\alpha}(\mathbf{r})\\\nonumber
\hat{\rho}_{c}(\mathbf{r})&=&e_0 \sum_{\alpha}^{J}\hat{\rho}_{\alpha}(\mathbf{r}) + {\rho}_{f}(\mathbf{r})
\end{eqnarray}
where $e_0$ is the electron charge and ${\rho}_{f}(\mathbf{r})$ represents the charge density of fixed charges in the system (by contrast to mobile ions).
In the following we will also use the inverse (in operator form) of the Coulomb potential. Since the Coulomb potential $v_c(\mathbf{r})$ in a dielectric medium of dielectric constant $\epsilon(\mathbf{r})$ satisfies the Poisson equation
\begin{equation}
\nabla_r \left( \epsilon(\mathbf{r})  \nabla_r  v_c(\mathbf{r}) \right) = -\delta(\mathbf{r}),
\end{equation}
it follows that its inverse operator is given by
\begin{eqnarray}
\label{inverse}
v_{c}^{-1}(\mathbf{r},\mathbf{r'})=-\nabla_{r} \left( \epsilon(\mathbf{r}).\nabla_{r'}\delta(\mathbf{r}-\mathbf{r'}) \right)
\end{eqnarray}

The following derivation can be easily generalized to more complex systems, e.g. where the solvent is not uniform, but made of freely orientable dipoles to represent water molecules, or even more refined ones\cite{Azuara:2008hb,Koehl:2010fi,Levy:2012gt,Buyukdagli:2016uk}. 
In the canonical ensemble, the Hamiltonian and partition function of the system are given in (\ref{defcanon}).
\begin{eqnarray}
\label{defcanon}
Z&=&\prod_{\alpha}\frac{1}{N_{\alpha}!}\prod \int d\mathbf{R}_{k}^{\alpha}\ e^{-\beta H}\\\nonumber
&H=&\frac{1}{2}\iint dr dr' \hat{\rho}_c(\mathbf{r})\ v_{c}(\mathbf{r}-\mathbf{r'})\ \hat{\rho}_c(\mathbf{r'})
\end{eqnarray}
Following steps described in Ref.~\citenum{Borukhov:2000gk,Netz:1999vd}, we build a field theory by replacing the integrals over the positions of the ions by integrals over density fields $n_{\alpha}$, by enforcing them in (\ref{defcanon}) with $\delta$-functions. Using a Fourier representation of these $\delta$-functions, we introduce the conjugate fields $\phi_{\alpha}(\mathbf{r})$. 
Shifting to the grand canonical ensemble, where each ion type $\alpha$ is now characterized by its fugacity $\lambda_{\alpha}$, one can rewrite the Hamiltonian and partition function of the system as in (\ref{defgrandcanon}).
\begin{eqnarray}\label{defgrandcanon}
\Xi&=&\int\prod_{\alpha}\mathcal{D}n_{\alpha}\mathcal{D}\phi_{\alpha}\ e^{-\beta \mathcal{H}_{\{n,\phi\}}}\\\nonumber
\beta\mathcal{H}_{\{n,\phi\}}&=&\frac{\beta}{2}\iint \rho_{c} v_{c}\rho_{c} -i \sum_{\alpha}\int n_{\alpha}\phi_{\alpha} - \sum_{\alpha}\lambda_{\alpha}\int  e^{-i\phi_{\alpha}}
\end{eqnarray}
where for the sake of simplicity, we have omitted the $dr$ in the integrals.

Due to the Gaussian nature of the partition function with respect to the density fields, it is possible to integrate it exactly w.r.t. these fields\cite{Borukhov:2000gk,Netz:1999vd}. Using the identity (\ref{inverse}), we obtain the well-known result \cite{Netz:2000kh}
\begin{eqnarray}
\label{potential}
\Xi&=&\int \mathcal{D}\phi \ e^{-\beta \mathcal{H}_{\{\phi\}}}\\\nonumber
\beta\mathcal{H}_{\{\phi\}}&=& \frac{\beta}{2}\int \epsilon (\mathbf{r}) \left( \nabla \phi(\mathbf{r})) \right)^2
 -i \beta \int {\rho}_{f}(\mathbf{r}) \phi(\mathbf{r}) - \sum_{\alpha}\lambda_{\alpha}\int  e^{-i\beta e_0 q_{\alpha}\phi(\mathbf{r})}
\end{eqnarray}

A powerful method to calculate the partition function (\ref{potential}) is through the saddle-point expansion (also called the loop expansion in quantum field theory). The idea is that the above integral is dominated by the extremum of the integrand (that is an extremum of the Hamiltonian) and thus one has to look for the saddle point of the Hamiltonian (\ref{potential}).
At the saddle-point level, the conjugate field is identified (up to a factor of $i$)  with the ‘physical’ electrostatic potential, solution to the PDE obtained by setting the functional derivative of the Hamiltonian to zero. This equation is the usual Poisson-Boltzmann equation.
\begin{eqnarray}\label{PBE}
\nabla\left(\epsilon({\bf r})\nabla\phi^{(0)} (\mathbf{r}) \right) = -\rho_{f}(\mathbf{r}) - \sum_{\alpha}q_{\alpha}e_{0}\lambda_{\alpha}e^{-\beta q_{\alpha}e_{0}\phi^{(0)}(\mathbf{r})}
\end{eqnarray}

The  integral which defines the partition function is approximated by the sole contribution from the saddle-point. It is thus a mean-field approximation. In certain cases, the contribution of the mean-field accounts well for the physics of the system. However, in many cases, in particular in complex fluids or biological systems, fluctuation effects can be important, and may even dominate the physics of these systems\cite{GronbechJensen:1997ej}.\\

The difficulty in sampling the partition function in (\ref{defgrandcanon}) or (\ref{potential}) is that the action (Hamiltonian) is complex. However, sampling requires positive Boltzmann weights (except for the complex Langevin method which we don't discuss here \cite{Ganesan:2001dz,Fredrickson:2002fo}). A way out of this difficulty is to write a representation of the partition function as a functional integral over the density fields only. In that case, as we shall see, the action becomes real, and thus the weights are positive. This can be achieved by integrating out the conjugate fields in (\ref{defgrandcanon}). 

Since the action in terms of the conjugate fields is local, the conjugate fields at different points are decoupled, and thus the integration over the conjugate fields can be performed independently at each point of space. The integration is performed by the saddle-point method (SPM) over the $\phi_{\alpha}$. The saddle point equations in  (\ref{defgrandcanon}) read
\begin{equation}
n_{\alpha}(\mathbf{r}) = \lambda_{\alpha} e^{-i \phi_{\alpha}^{(0)}(\mathbf{r})}
\end{equation}
which implies $i\phi_{\alpha}^{(0)}=-\log{\frac{n_{\alpha}}{\lambda_{\alpha}}}$, thus yielding an expression of the  Hamiltonian in terms of the fluctuating density fields only
\begin{eqnarray}\label{defhamil}
\Xi&=&\int\prod_{\alpha}\mathcal{D}n_{\alpha}\ e^{-\beta \mathcal{H}_{\{n\}}}\\\nonumber
\beta\mathcal{H}_{\{n\}}&=&\frac{\beta}{2}\iint \rho_{c} v_{c}\rho_{c} + \sum_{\alpha}\int n_{\alpha}\log{\frac{n_{\alpha}}{\lambda_{\alpha}e}} 
\end{eqnarray}
where the functional integral over the density fields is only over {\em positive fields}.
Note that this expression of the Hamiltonian is identical to that used in density functional theory of fluids \cite{Jeanmairet:2015uj,Sushko:2015vl}. We have chosen to perform the $\phi_{\alpha}$ integral at saddle-point level only, but it is clear that one could perform the saddle-point expansion to higher order\cite{Netz:2000kh}. 
\\
Since the densities are positive and the Hamiltonian is real, it is now possible to perform stochastic sampling on (\ref{defhamil})\cite{Negele:1988uh}. 
Although it is possible to use a Monte Carlo method, we will sample using a Langevin equation approach, as it allows for additional flexibility.

%%%%%%%%%%%%%%%%%%%%%%%%%%%%%%%%%%%%%%%%%%%%%%%%%%%%%%%
%We start by shifting the Hamiltonian around the mean-field solution (Eq.\ref{defshiftedhamil}). It should be noted that the saddle-point on the podensity corresponds to the mean-field solution in the potential description.
%\begin{eqnarray}\label{defshiftedhamil}
%\beta\mathcal{H}&=&\beta\mathcal{H}_{0}+\beta\mathcal{H}_{u}\\\nonumber
%&& \text{where } \beta\mathcal{H}_{0}=\frac{\beta}{2}\iint \rho_{c}^{(0)} v_{c}\rho_{c}^{(0)} + \sum_{\alpha}\int n_{\alpha}^{(0)}\log{\frac{n_{\alpha}^{(0)}}{\lambda_{\alpha}e}}\\\nonumber
%&&\text{and } \beta\mathcal{H}_{u}=\frac{\beta}{2}\iint \rho_{c}^{(u)} v_{c}\rho_{c}^{(u)} + \sum_{\alpha}\int\Big[ (n_{\alpha}^{(0)}+u_{\alpha})\log\Big({1+\frac{u_{\alpha}}{n_{\alpha}^{(0)}}}\Big)-u_{\alpha}\Big]\\\nonumber
%&&\text{with }n_{\alpha} = n_{\alpha}^{(0)} + u_{\alpha}
%\end{eqnarray}
The general form of the overdamped Langevin equation is a stochastic diffusion equation, where the stochastic force term depends on the diffusion tensor in order to ensure detailed balance\cite{Negele:1988uh}
\begin{eqnarray}\label{spde}
\dot{n}_{\alpha}(\mathbf{r},t)&=&-\int D(\mathbf{r},\mathbf{r'})\frac{\partial \beta \mathcal{H}_{\{n\}} }{\partial n_{\alpha}(\mathbf{r'},t)}+\xi_{\alpha}(\mathbf{r},t)\\\nonumber
<\xi_{\alpha}(\mathbf{r},t)\xi_{\alpha'}(\mathbf{r'},t')>&=&2D(\mathbf{r},\mathbf{r'})\delta(t-t')\delta_{\alpha,\alpha'}
\end{eqnarray}

Any definite positive diffusion tensor $D(\mathbf{r},\mathbf{r'})$ in (\ref{spde}) guarantees detailed balance, and thus correct sampling of the partition function (\ref{defhamil}).

As is well-known, the choice of the diffusion tensor affects the dynamics of the fluctuating fields, and their conservation. The most well-known examples are the models A and B of Hohenberg and Halperin\cite{Hohenberg:1977fq}, where the latter is conservative with respect to the fluctuating field, while the former is not. Some authors have proposed other stochastic equations to reflect the underlying dynamics at the particle level. This dynamics bears some similarity to Model B, with the major difference that the noise is multiplicative with respect to the fluctuating density\cite{Kawasaki:1966dp,Dean:1996df}.\\
Here we do not aim at describing the transport processes or the system kinetics, and our only requirement is that we sample the partition function with the proper weights, in order to ensure detailed balance.\\
If we choose a diffusion tensor proportional to the Dirac delta function (diffusion scalar) 
$D(\mathbf{r},\mathbf{r'})= D_0 \epsilon(\mathbf{r})\delta(\mathbf{r}-\mathbf{r'})$, the Langevin equation becomes 
\begin{eqnarray}
\label{modelA}
\dot{n}_{\alpha}(\mathbf{r},t)&=&- D_0 \Bigg(\log \frac{n_{\alpha}(\mathbf{r},t)}{\lambda_{\alpha}} + \beta q_{\alpha} e_0^2 \int dr' v_c(\mathbf{r}-\mathbf{r'}) \rho_{c}(\mathbf{r'},t) \Bigg) +\xi_{\alpha}(\mathbf{r},t)\\\nonumber
&&<\xi_{\alpha}(\mathbf{r},t)\xi_{\alpha'}(\mathbf{r'},t')>=2D_0 \epsilon(\mathbf{r})\delta(\mathbf{r}-\mathbf{r'})\delta(t-t')\delta_{\alpha,\alpha'}
\end{eqnarray}

This Langevin equation is very straightforward to solve numerically. However, it involves the long range potential $v_c$ and thus each update of the density fields requires an integral over all space.

Note that when $n_{\alpha}$ goes to zero in eq.(\ref{modelA}), the entropic force $-\log \frac{n_{\alpha}}{\lambda_{\alpha}}$ becomes infinitely repulsive, thus preventing 
the densities from becoming negative. However in practice, this repulsive force is quite small, and thus one needs special care to ensure the positivity of the densities.

The equation (\ref{spde}) can be made local in space by a specific choice of the diffusion tensor, which is identical to model B of Ref. \citenum{Hohenberg:1977fq}. Taking this tensor to be proportional to the inverse of the Coulomb operator (which is definite positive as its Fourier transform is proportional to $1/k^2$), $D(\mathbf{r},\mathbf{r'})=D_{0}v_{c}^{-1}(\mathbf{r}-\mathbf{r'})$,
eq. (\ref{spde}) becomes local and takes the form
\begin{eqnarray}
\label{localspde}
\dot{n}_{\alpha}(\mathbf{r},t)&=&D_{0}\Big[\nabla \big[\epsilon(\mathbf{r})\nabla\log \frac{n_{\alpha}(\mathbf{r},t)}{\lambda_\alpha}\big]-\beta e_{0}^{2}q_{\alpha}\sum_{\alpha'}q_{\alpha'}n_{\alpha'}(\mathbf{r},t)\Big]+\nabla.\xi_{\alpha}(\mathbf{r},t)
\end{eqnarray}
where $\xi_{\alpha}$ is a Gaussian noise satisfying
\begin{eqnarray}
\label{noise}
\langle \xi_{\alpha}(\mathbf{r},t) \rangle &=&0 \nonumber  \\
\langle \xi_{\alpha}(\mathbf{r},t)  \xi_{\alpha'}(\mathbf{r'},t') \rangle &=& 2 D_0 \epsilon(\mathbf{r}) \delta _{\alpha,\alpha'} \delta (\mathbf{r}-\mathbf{r'}) \delta(t-t')
\end{eqnarray}

It is important to note at this stage that all the physical effects accounted for by generalizations of the Poisson-Boltzmann approach (mixtures of ions with difference sizes, non-uniform polarizable medium, and steric exclusion - to name a few)
can be implemented in the present approach, by using a more general form for the charge densities.
 
When solving numerically the above equation, the problem of the positivity of $n_{\alpha}$ arises, and in order to overcome it as efficiently as possible, it is useful to work on the density fluctuation $u_{\alpha}$ around the static mean-field $n_{\alpha}^{(0)}$
\begin{equation}
n_{\alpha}=n_{\alpha}^{(0)} + u_{\alpha}
\end{equation}
where the static mean-field $n_{\alpha}^{(0)}$ is defined by the  equation
\begin{equation}
\label{PBdif}
\nabla\big[\epsilon(\mathbf{r})\nabla\log \frac{n_{\alpha}^{(0)}(\mathbf{r})}{\lambda_\alpha}\big]=\beta e_{0}^{2}q_{\alpha}\sum_{\alpha'}q_{\alpha'}n_{\alpha'}^{(0)}(\mathbf{r})
\end{equation}
which can be easily shown to be identical to the Poisson-Boltzmann equation.

It should be noted that shifting the density around the mean-field solution allows one to rewrite the Hamiltonian of the system as two additive terms, one corresponding to the free energy of the system in the mean-field approximation, and the other one accounting for the "instantaneous" fluctuations, which correct for the former when properly integrated.
\begin{eqnarray}\label{defshiftedhamil}
\beta\mathcal{H}&=&\beta\mathcal{H}_{0}+\beta\mathcal{H}_{u}\\\nonumber
\text{where } \ \ \beta\mathcal{H}_{0}&=&\frac{\beta}{2}\iint \rho_{c}^{(0)} v_{c}\rho_{c}^{(0)} + \sum_{\alpha}\int n_{\alpha}^{(0)}\log{\frac{n_{\alpha}^{(0)}}{\lambda_{\alpha}e}}\\\nonumber
\text{and } \ \ \beta\mathcal{H}_{u}&=&\frac{\beta}{2}\iint \rho_{c}^{(u)} v_{c}\rho_{c}^{(u)} + \sum_{\alpha}\int\Big[ (n_{\alpha}^{(0)}+u_{\alpha})\log\Big({1+\frac{u_{\alpha}}{n_{\alpha}^{(0)}}}\Big)-u_{\alpha}\Big]
\end{eqnarray}
The local Langevin equation for the fluctuation field becomes then
\begin{equation}
\label{flucspde}
\dot{u}_{\alpha}(\mathbf{r},t)=D_{0}\Big[\nabla\big[\epsilon(\mathbf{r})\nabla\log (1+\frac{u_{\alpha}(\mathbf{r},t)}{n_{\alpha}^{(0)}(\mathbf{r})}) \big]-\beta e_{0}^{2}q_{\alpha}\sum_{\alpha'}q_{\alpha'}u_{\alpha'}(\mathbf{r},t)\Big]+\nabla.\xi_{\alpha}(\mathbf{r},t)
\end{equation}
where the Gaussian white noise $\xi_{\alpha}$ is defined by eq. (\ref{noise}). 

In the following, we show how this local PDE can be solved numerically and discuss some examples.
%%%%%%%%%%%%%%%%%%%%%%%%%%%%%%%%%%%%%%%%%%%%%%%%%%%%
\section{Implementation}

The solver for the SPDE derived above has been implemented by starting from AquaSol, a solver for the generalized Poisson-Boltzmann-Langevin equation\cite{Koehl:2010fi}. We describe here the time and space discretization.

\subsection{Time discretization and non-negativity enforcing}

Starting from the solution to the Poisson-Boltzmann equation (\ref{PBE}), the initial configuration for the fluctuating field is set to zero everywhere. Time discretization is performed using a simple first-order Euler method. The time index is denoted by $i$, the time-step by $\Delta t$, the drift term by $F_{1}$ and the diffusive term by $F_{2}$
\begin{eqnarray}
u_{\alpha}(\mathbf{r},i+1) &=& u_{\alpha}(\mathbf{r},i) + D_{0}\Delta t F_{\alpha}^{(1)}(\mathbf{r},i) + \sqrt{2D_{0}\Delta t}F_{\alpha}^{(2)}(\mathbf{r},i)\\ \nonumber
\text{where} \ \ F_{\alpha}^{(1)}(\mathbf{r},i) &=&\nabla . \left(\epsilon(\mathbf{r})\nabla S_{\alpha}(\mathbf{r},i)\right)  - \beta e_{0}^2 q_{\alpha}\sum_{\alpha'}q_{\alpha'}u_{\alpha'}(\mathbf{r},i)\\\nonumber
F_{\alpha}^{(2)}(\mathbf{r},i)&=&\nabla .\left(\sqrt{\epsilon(\mathbf{r})} \zeta_{\alpha}(\mathbf{r},i)\right)
\end{eqnarray}
with
\begin{eqnarray}
\label{noise}  
<\zeta_{\alpha}^{(x)}(\mathbf{r},i)\zeta_{\alpha'}^{(x')}(\mathbf{r'},i')>&=&\delta_{\alpha,\alpha'}\delta_{i,i'}\delta_{x,x'}\delta(\mathbf{r}-\mathbf{r'}) \\
\text{and}\ \  S_{\alpha}(\mathbf{r},i) &=& \log\Big(1+\frac{u_{\alpha}(\mathbf{r},i)}{n_{\alpha}^{(0)}(\mathbf{r})}\Big) \nonumber
\end{eqnarray}

\begin{figure}[H]
    \centering
    \includegraphics[width=\textwidth]{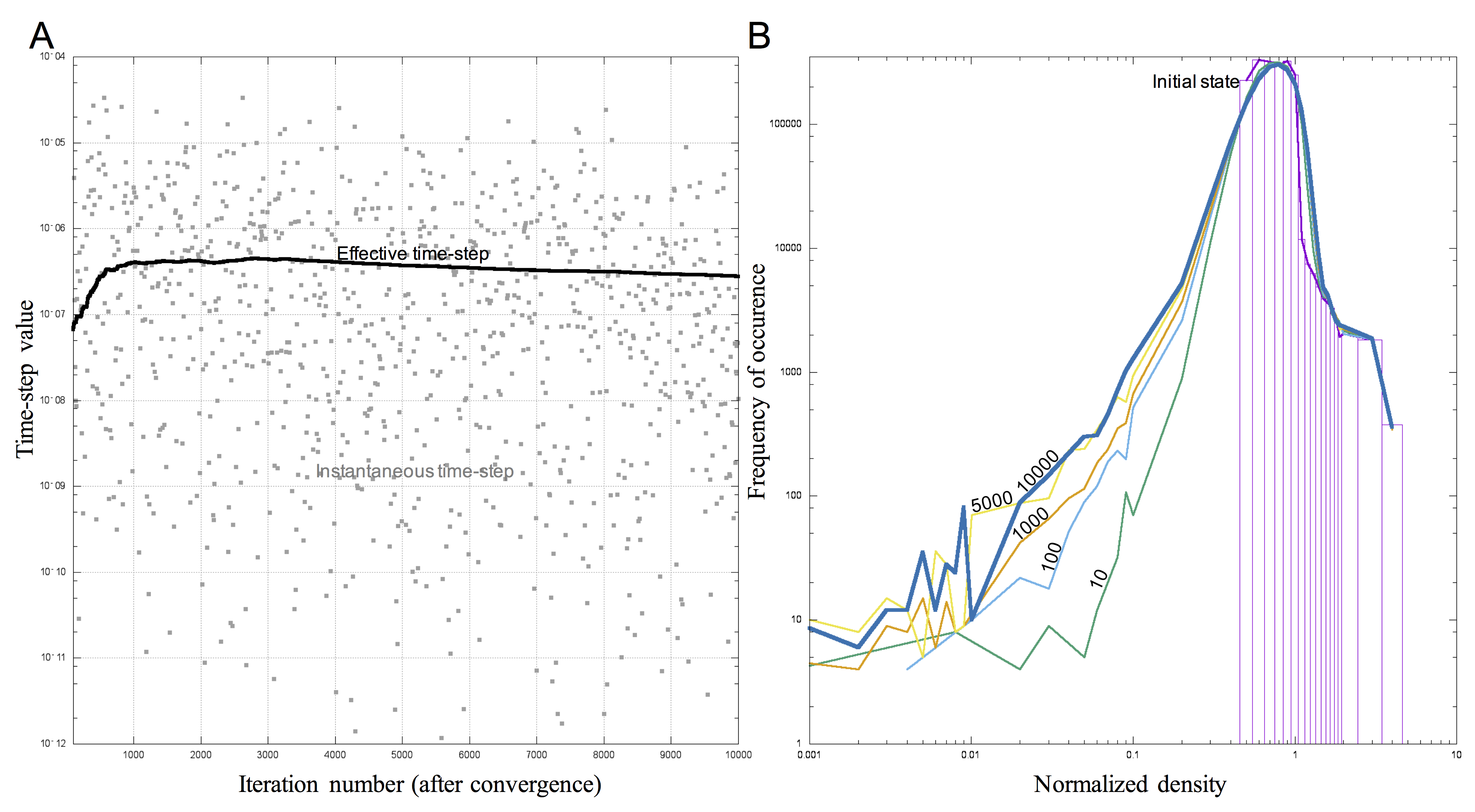}
    \caption{\footnotesize Convergence of the time-step and density distribution illustrated on a typical simulation run for one of the systems detailed in the Results section. (A) Adaptive time-stepping. The time-step $\Delta$t is plotted as a function of the iteration number. The "instantaneous" time-step, adapted to guarantee non-negativity of the density everywhere on the grid at the next step, is shown as a grey dot. The "effective" time-step, defined as the cumulated time divided by the iteration number, is plotted as a black line. (B) Density distribution through the simulation (log-log scale). Starting from the Poisson-Boltzmann density distribution, displayed as purple boxes, the density distribution is plotted as colored lines at the following iteration number: 10(green), 100(blue), 1000(orange), 5000(yellow) and 10000(thick blue). It can be seen that following a rapid diffusion toward both high and low density values, the system converges after a few thousands steps. }
    \label{figcvrgce}
\end{figure}

As mentioned earlier, care must be taken as to how the positivity of the density is enforced. In fact, in its continuous formulation, this property is guaranteed by construction, since there is a divergent repulsive force at zero densities. However, the presence of a large stochastic term due to the time discretization scheme may allow the densities to jump from a small positive value to a small negative value in a single timestep. Among the possible strategies, we implemented a scheme where when one density becomes negative, we go one step back and rescale the time step iteratively until the density at any point in the box will not drop below zero. In practice, we checked that the average time-step reaches a stable value after a transient equilibration period (See Fig.\ref{figcvrgce}-A).

\subsection{Space discretization}

We define a cubic volume discretized as a uniform grid of step size $a$, with the same number N of points in each direction. The coordinates of any point of the grid is indexed with $h,k,l$ integers, each one running from 1 to N. For consistency with the PBE solver of AquaSol\cite{Koehl:2010fi}, the finite volume method (FVM) will be used to discretize the SPDE. This method does not discretize directly the continuous function (as in a Finite Difference method) but rather discretizes space and considers discrete functions that are averages of the continuous function within each elementary voxel $\Omega_{hkl}$. However, we will abusively keep the same notation for the original function and its discrete average (Eq.\ref{fvmall}). In the following, the lattice spacing is denoted by $a$. The discretized  form of the noise , in particular of the spatial delta function appearing in the correlation function Eq. (\ref{noise}), yields the correlation function
\begin{equation}
<\zeta_{\alpha}^{(x)}(\Omega_{hkl},i)\zeta_{\alpha'}^{(x')}(\Omega_{h'k'l'},i')>=\frac{1}{a^3} \delta_{\alpha,\alpha'}\delta_{i,i'}\delta_{x,x'}\delta_{hkl,h'k'l'}
\end{equation}
which suggests to rescale the Gaussian $\zeta$ by a factor $\sqrt{a^{3}}$.
The drift term is the sum of an entropic and a Coulomb term. The latter, being local, is dealt with in a straightforward manner. The former is explicited in Eq.\ref{fvmf1} using the divergence theorem. %However, care must be taken to preserve the statistics while summing. 
Denoting by $\Gamma_{hkl}$ the surface of the voxel $\Omega_{hkl}$, and by $d\boldsymbol\Gamma$ the infinitesimal surface element, we have
\begin{eqnarray}\label{fvmall}
u_{\alpha}(\mathbf{r}_{hkl}, i+1) &=& u_{\alpha}(\mathbf{r}_{hkl}, i) \\\nonumber
&+& a^{-3}\Big\{D_{0}\Delta t\int_{\Omega_{hkl}}d\mathbf{r}F_{\alpha}^{(1)}(\mathbf{r},i) + \sqrt{2D_{0}\Delta t}\int_{\Omega_{hkl}}d\mathbf{r}F_{\alpha}^{(2)}(\mathbf{r},i)\Big\}
\end{eqnarray}
\begin{align}\label{fvmf1}
\int_{\Omega_{hkl}} d\mathbf{r}\ \nabla.\Big\{\epsilon(\mathbf{r})\nabla S_{\alpha}(\mathbf{r},i)\Big\} &=
	& \oint_{\Gamma_{hkl}}
		&\epsilon(\mathbf{r})\nabla S_{\alpha}(\mathbf{r},i).d\boldsymbol\Gamma\\\nonumber
&\sim
	& a^{2}\sum_{u=h,k,l} \Big[
		&\epsilon\big(\mathbf{r}_{hkl}+\frac{1}{2}\mathbf{e}_{u}\big)\nabla S_{\alpha}\big(\mathbf{r}_{hkl}+\frac{1}{2}\mathbf{e}_{u},i\big)\\\nonumber
&
	&
		&-\epsilon\big(\mathbf{r}_{hkl}-\frac{1}{2}\mathbf{e}_{u}\big)\nabla S_{\alpha}\big(\mathbf{r}_{hkl}-\frac{1}{2}\mathbf{e}_{u},i\big)
		\Big]\\\nonumber
&\sim
	& a^{2}\sum_{u=h,k,l} \Big[
		&\Big(\frac{\epsilon\big(\mathbf{r}_{hkl}+\frac{1}{2}\mathbf{e}_{u}\big)}{a}\Big)\\\nonumber
&
	&
		&\ \ \ \times\Big(S_{\alpha}(\mathbf{r}_{hkl}+\mathbf{e}_{u},i)-S_{\alpha}(\mathbf{r}_{hkl},i)\Big)\\\nonumber
&
	& 
		&+\Big(\frac{\epsilon\big(\mathbf{r}_{hkl}-\frac{1}{2}\mathbf{e}_{u}\big)}{a}\Big)\\\nonumber
&
	&
		&\ \ \ \times\Big(S_{\alpha}(\mathbf{r}_{hkl}-\mathbf{e}_{u},i)-S_{\alpha}(\mathbf{r}_{hkl},i)\Big)
		\Big]
\end{align}

Similarly, the diffusive term can be cast as a surface integral around each voxel 
\begin{align}\label{fvmf2}
\int_{\Omega_{hkl}} d\mathbf{r}\ \nabla \left(\sqrt{\epsilon(\mathbf{r})}\boldsymbol\zeta_{\alpha}(\mathbf{r},i)\right) &= \oint_{\Gamma_{hkl}} \sqrt{\epsilon(\mathbf{r})}\boldsymbol\zeta_{\alpha}(\mathbf{r},i).d\boldsymbol\Gamma\\\nonumber
&\sim\ 
a^{-3/2}\sum_{u=h,k,l}a\Bigg[
\sqrt{\epsilon(\mathbf{r}_{hkl}+\frac{1}{2}\mathbf{e}_{u})}\ \zeta_{\alpha}^{(u)}(\mathbf{r}_{hkl}+\frac{1}{2}\mathbf{e}_{u},i)\\\nonumber
&-\sqrt{\epsilon(\mathbf{r}_{hkl}-\frac{1}{2}\mathbf{e}_{u})}\ \zeta_{\alpha}^{(u)}(\mathbf{r}_{hkl}-\frac{1}{2}\mathbf{e}_{u},i)\ \ \Bigg]
\end{align}
where $\mathbf{e}_{u}$ is the lattice vector in direction $u$ and the correlation function of the Gaussian noise is given by
\begin{equation}
<\zeta_{\alpha}^{(x)}(\mathbf{r}_{hkl},i)\zeta_{\alpha'}^{(x')}(\mathbf{r'}_{h'k'l'},i')>=\delta_{\alpha,\alpha'}\delta_{i,i'}\delta_{x,x'}\delta_{hkl,h'k'l'}
\end{equation}

In order to prevent stability problems as the size $a$ decreases, we rescale the diffusion scalar $D_{0}$ with the voxel volume. After introducing the following notations 
 $D_{d}=D_{0}\epsilon_{0}a^{3}\Delta t$ and $l_{B}=\frac{\beta e_{0}^{2}}{4 \pi \epsilon_{0}}$, the SPDE can be rewritten in a compact form as in Eq.\ref{eqdiscrete}.
\begin{eqnarray}\label{eqdiscrete}
u_{\alpha}(hkl,i+1) &=& u_{\alpha}(hkl,i) \\\nonumber
&+& a^{-3}\Bigg(D_{d}\Big\{\sum_{\pm}\sum_{v=h,k,l}A(hkl \pm v)E_{\alpha}(hkl\pm v,i) - H_{\alpha}(hkl,i)\Big\}\\\nonumber
&&+ \sqrt{2D_{d}}\Big\{\sum_{\pm}\sum_{v=h,k,l}\pm A^{1/2}(hkl\pm v)\zeta_{\alpha}^{(v)}(hkl\pm v,i)\Big\}\Bigg)\\\nonumber
\text{where } && A(hkl\pm v) = \frac{a}{\epsilon_{0}}\epsilon(\mathbf{r}_{hkl}\pm\frac{1}{2} \mathbf{e}_{u})\\\nonumber
&& E_{\alpha}(hkl\pm v,i)=S_{\alpha}(\mathbf{r}_{hkl}\pm \mathbf{e}_{v},i)-S_{\alpha}(\mathbf{r}_{hkl},i)\\\nonumber
&& H_{\alpha}(hkl,i) = 4\pi l_{B}a^{3}q_{\alpha}\sum_{\alpha'}q_{\alpha'}u_{\alpha'}(\mathbf{r}_{hkl},i)
\end{eqnarray}

To fully define our problem, the SPDE just described must be complemented with boundary conditions. The boundary conditions are dictated by the ones used to solve the mean-field problem. As periodic conditions are not yet implemented in AquaSol\cite{Koehl:2010fi} , we only implemented the equivalent of the Neumann boundary condition, i.e. that any incoming flux is set to zero: a surface term for a voxel on the box boundary is set to zero.\\
Finally, the integration of the SPDE must reach a stationary sampling regime in the long time limit. Convergence is monitored by measuring the standard deviation of the density distribution over the grid as a function of the iteration number. Once its value over a given number of iterations does not change by more than a given threshold, the system is considered to have reached convergence. It is illustrated in Figure \ref{figcvrgce}-B where the total density distribution is seen to reach a stationary regime after a few thousand steps. When observables need to be averaged, the actual averaging only starts once convergence has been reached.

\section{Results and Discussion}

\begin{figure}[H]
    \centering
    \includegraphics[width=0.75\textwidth]{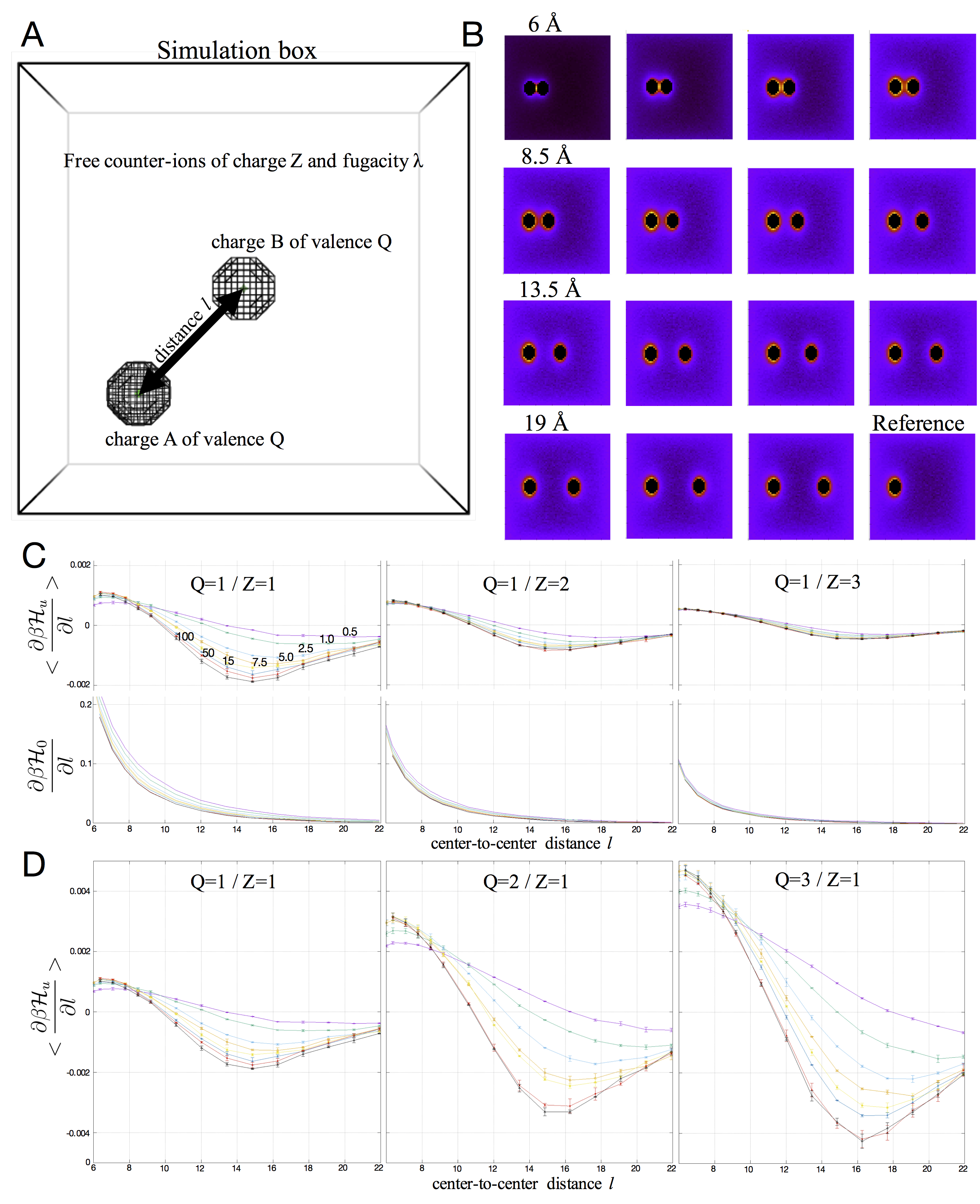}
    \caption{\footnotesize Like-charge attraction. (A) Illustration of the simulation cubic box,  where each side is 32 \AA long with 65 equally spaced points. The solvent-excluded volume of the two 2.4 \AA\ radius spheres is shown as a mesh for one configuration of the system where the "moving" sphere is found at the center of the box. (B) Illustration as a heat map of the counter-ion density in the plane containing the two like-charge ions at increasing distance $l$ from each other. (C) The two components of the force between the two like-charge ions of valence 1, surrounded by counter-ions of valence 1 to 3 (left to right) and increasing fugacity (0.5M to 100M as indicated on the left inset) are represented: the mean-field repulsive (positive) force is displayed at the bottom, while the contribution arising from the counter-ion density fluctuations is plotted at the top. (D) Same illustration as in C, only this time for systems of constant counter-ion valence of 1, and increasing like-charge ion valences from 1 to 3 (left to right).}
    \label{figresults}
\end{figure}

In the following, we consider two spheres of same radius  with a point-charge of valence $Q$ at their center, immersed in a cubic box of finite size containing counter-ions of valence $Z$ with fugacity $\lambda$ (see Fig.\ref{figresults}-A). One of the charges is fixed while the other one is moving along the axis defined by the unit vector $\hat{\mathbf{I}}$ that joins them.\\
The counter-ions are free to move in the uniform dielectric background of the box, although their density is constrained to zero in the solvent-excluded volume defined by the two fixed spheres, and is constrained by the boundary conditions at the edges of the box.\\
We want to measure the force between the two fixed spherical ions as a function of their separation $l$. The free energy $\mathcal{F}(l)$ of the system as a function of $l$ is given by
\begin{eqnarray}
\beta \mathcal{F}(l) &=& -\log\text{Tr}e^{-\beta \mathcal{H}(l)} \nonumber \\
&=&  \beta \mathcal{H}_{0}(l) -\log\text{Tr}e^{-\beta \mathcal{H}_{u}(l)}
\end{eqnarray}
and the effective force $\mathbf{F(l)}=F(l)\ \hat{\mathbf{I}}$ between the two charges is the so-called "potential of mean force", which is the derivative of the free energy with respect to the separation
\begin{eqnarray}
F(l)&=& - \frac{\partial \mathcal{F}}{\partial l}  \nonumber \\
&=&- \langle \frac{\partial \mathcal{H}}{\partial l} \rangle \nonumber \\
&=&  - \frac{\partial \mathcal{H}_{0}}{\partial l} - \langle \frac{\partial \mathcal{H}_{u}}{\partial l} \rangle
\end{eqnarray}
where the brackets $\langle \ldots \rangle$ denote the average over the Boltzmann weight $\text{Tr}e^{-\beta \mathcal{H}_{u}(l)}$.
We see that the force between the charges can be decomposed into a mean-field and a fluctuation component. The mean-field part reduces to the electric  field $\mathbf{E}_{0}$ exerted by one charge on the other at the mean-field level. The total force is thus given by
\begin{eqnarray}
\label{meanforce_theory}
\mathbf{F(l)} &=& Q\mathbf{E}_{0} -\frac{1}{\beta} \int\frac{\partial n^{(0)}(\mathbf{r})}{\partial l}\langle\log\Big(1+\frac{u(\mathbf{r})}{n^{(0)}(\mathbf{r})}\Big)-\frac{u(\mathbf{r})}{n^{(0)}(\mathbf{r})}\rangle \hat {\mathbf{I}}
\end{eqnarray}

In practice, considering two fixed charges at distance $l$ in a given solution, Eq.\ref{meanforce_theory} requires the knowledge of the mean-field electric field $\mathbf{E}_{0}$ on one charge, the differential mean-field ion densities $\frac{\partial n^{(0)}(\mathbf{r})}{\partial l}$ and the time-average $\langle...\rangle$ of the term depending on the fluctuating densities. As one charge (B) is displaced further away from the other (A) (see Fig.\ref{figresults}-B for an illustration), the mean-field electric field is measured on (A) $via$ a centered finite difference scheme using the value of the mean-field electrostatic potential on the grid points around it. The differential mean-field ion densities are obtained through slight symmetric perturbation $\delta l$ of charge (A) in the direction of charge (B) - of the order of a hundredth of an Angstrom - and subsequent centered finite differentiation of the perturbed densities. The time-average of the latter component in Eq.\ref{meanforce_theory} is accumulated as Eq.\ref{eqdiscrete} is integrated - using the adaptative time-stepping scheme.\\
In order to get rid of components of the force that arise from the finite size of the box and the non-periodic treatment of the boundaries, from the force at distance $l$, we subtract the force computed 
when only charge (A) is present (see Fig.\ref{figresults}-B, bottom right map). This latter situation representing a proxy for the infinite distance separation configuration.\\

For our illustrative purpose, we consider a cubic box with $N=65$ and $a=0.5\ \angstrom$, with two fixed ions located on the diagonal of the mid z-plane (see Fig.\ref{figresults}). Their valence was set to -1, -2 or -3, and a constant radius of 1$\ \angstrom$, which added to the probe radius of $1.4\ \angstrom$ defines an excluded quasi-spherical volume with dielectric permittivity constant set to 1. Outside this volume, the permittivity constant is set to 80. We consider a system with counter-ions only in the solvent volume. Their valence ranges from 1 to 3, and their fugacity from 0.5 M to 100 M.
For each condition, the simulation was repeated 10 times with different seed values to initiate the random number generator, allowing to retrieve the average force profile together with the associated error.\\

The overall shape of the profile of both the mean-field component of the force between the like-charges and the one arising from the density fluctuations, seems conserved within the parameter space explored. The mean-field component (Fig.\ref{figresults}.C-bottom) is always repulsive (positive) and decreases monotonically with the separation distance $l$ between the two charges. The correction term has a more elaborate shape: it is repulsive at short distance and becomes attractive as the distance increases, reaching a minimum after which the profile asymptotically goes back to zero.\\
Despite an overall conservation of the profile's shape, the relative amplitude of the features just described varies as a function of the ion's valence and fugacity (i.e. as the coupling varies).\\
As expected, the screening effect of the mean-field component of the force increases as a function of both the counter-ion fugacity and valence (Fig.\ref{figresults}.C-bottom). The component arising from the density fluctuations is several order of magnitude below that of the mean-field component at short distance, but their ratio increases in an intermediate range (Fig.\ref{figresults}.C-top). In the conditions shown in Fig.\ref{figresults}.C however, the magnitude of the fluctuation-induced component does not increase enough to compensate the repulsive mean-field component.\\
Interestingly, as can be seen on Fig.\ref{figresults}.D, the force arising from the counter-ion fluctuations increases with the like-charge valency. The mean-field component is not shown however, as the finite-difference scheme used to measure it on the grid becomes flawed with numerical inaccuracy as the valence of the ion is increased. This is one of the limitations of our current implementation of the method. Indeed, in order to maintain the same level of precision when computing the electric field on a charge of increasing valence, using a finite difference approach requires an appropriate rescaling of the grid size $a$, and thus a rapid increase of the number of points within the grid. Although not impossible in theory, this approach quickly proves impractical, and in fact calls for a better devised space discretization approach. We aim to implement the finite-element method\cite{Holst:2012ab} in AquaSol\cite{Koehl:2010fi} in a near future to solve this precision issue with respect to the mean-field electric field, as well as because such discretization schemes have been shown to be more reliable when solving SPDE\cite{delaTorre:2015ej}.\\
Despite this limitation, it should be noted that, in the distance range where the fluctuations are maximally attractive, the mean-field component decreases as the valency of the fixed charges increases, leading to a situation where the fluctuating force might become strong enough to overcome the repulsive mean-field component, thus potentially resulting in an effective like-charge attraction\cite{GronbechJensen:1997ej,Naji:2005ul}. We let for future work the detailed characterization of this effect as the parameter space is systematically explored.

Our approach and its implementation presented here, represents one further step towards the development of a performing computational tool for the quantitative treatment of Coulomb interactions in systems of arbitrary geometry. We believe that such a tool will find many applications in many different fields, such as structural molecular biology\cite{Qiu:2010ja,Anthony:2012kj,Sauguet:2013gj,Gillespie:2014kk}, where the study of the dynamics of large polyelectrolyte  systems in complex solvents is still in demand for more efficient and accurate methods\cite{Koehl:2014cr}.

%%%%%%%%%%%%%%%%%%%%%%%%%%%%%%%%%%%%%%%%%%%%%%%%%%
\begin{acknowledgement}

The authors thank Patrice Koehl for continuous discussion and support. FP gratefully acknowledges support from ANR blanc grant "Fluctuations in Structured Coulomb Fluids".

\end{acknowledgement}

%\begin{suppinfo}
%
%A listing of the contents of each file supplied as Supporting Information
%should be included. For instructions on what should be included in the
%Supporting Information as well as how to prepare this material for
%publications, refer to the journal's Instructions for Authors.
%
%The following files are available free of charge.
%\begin{itemize}
 % \item Filename: brief description
 % \item Filename: brief description
%\end{itemize}
%
%\end{suppinfo}

%%%%%%%%%%%%%%%%%%%%%%%%%%%%%%%%%%%%%%%%%%%%%%%%%%
\bibliography{coulomb_spde}

\end{document}